\documentclass[twocolumn,showpacs,amsmath,amssymb]{revtex4}

\usepackage{epsfig}
           
\usepackage{dcolumn}

\usepackage{bm}

\usepackage{amsmath}

\usepackage{amsfonts}

\usepackage{amssymb}

\usepackage{graphicx}%

\newcommand{\beq}{\begin{eqnarray}}

\newcommand{\eeq}{\end{eqnarray}}

\begin{document}

\title{Symmetry and environment effects  on rectification mechanisms in quantum pumps}

\author{Liliana Arrachea }

\affiliation{Instituto de Biocomputaci\'on y F\'{\i}sica de Sistemas
  Complejos, Universidad de Zaragoza,
Corona de Arag\'on 42, (50009) Zaragoza, Spain.
}


\begin{abstract}
We consider a paradigmatic model of quantum pumps and discuss its rectification
properties in the framework of a symmetry analysis proposed for ratchet systems.
We discuss the role of the environment in breaking time-reversal symmetry
and the possibility of a finite directed current in the Hamiltonian limit
of annular systems.
\end{abstract}

\maketitle

{\em Introduction}.
Recently there has been  a good amount of experimental and theoretical 
activity devoted to study quantum pumps \cite{swi,qupum,qupumpt,mosk1,mosk2}
and quantum ratchets \cite{qurat}. The basic underlying idea  is
the generation of  a net current as a response to a time-dependent external
field without a net static bias. The potential applications of this effect 
captures increasing interest within the communities of condensed matter
physics and chemistry. 
 
The paradigm of a ratchet system is a device with broken spacial symmetry
affected by a zero-mean time-dependent force. An additional ingredient is the
  coupling to an
environment, which is  usually  represented by reservoirs or some external noise.
An important point in the investigation of the ratchet effect
has been the understanding of the role played by the symmetries in the rectification
properties of the related devices. In particular, a very simple criterion 
 has been proposed in order to decide  whether a system 
driven by a time-dependent field is able to support a dc-current
\cite{ser}. It could be stated as follows: current rectification is not
possible in systems  where the symmetry operations leading to a change in
the sign of the relevant current, $J(t) \rightarrow - J(t)$,
 leave the ensuing equations for its time evolution  invariant.
  This criterion is completely general and applies to both adiabatic and
  non-adiabatic regimes. Its validity has been mainly explored in the
  framework of classical systems.

Recent advances of material science, have enabled the experimental realization of the
ratchet effect in quantum pumps \cite{swi,qupum}. Charge and spin currents have been generated
as a response to two harmonic potentials  with a phase lag, which
induce out-of-phase
oscillations at the walls of a quantum dot. Experiments have been performed in linear
arrays where the quantum dot is in contact to leads. Under these
operational conditions these devices are actually open quantum systems and
time-inversion symmetry
breaking is introduced in the problem not only because of the phase lag between the
potentials but also through the coupling to the environment. An interesting
alternative setup is obtained by bending the structure to form a ring to generate
 a dc-current along its circumference. An important example of this class
 is a ring threaded by a time-dependent magnetic 
flux. In the case of a flux with a linear dependence on time, Bloch oscillations
take place due to the induced constant electric field and
the coupling to reservoirs is essential to rectify this current
\cite{ringo,lilir}.

\begin{figure}
\centerline{\psfig{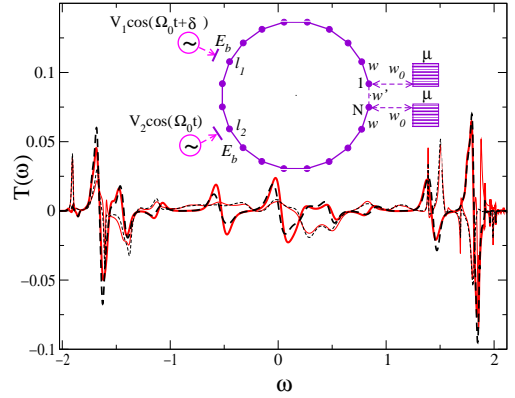}}
\hfill
\caption{\small (Color online) Transmission function for a ring with $N=20$ sites,
$l_1=9,l_2=12$  barriers of height $E_b=1$, $V_1=V_2=0.5$, $w_0=1$ and $\delta=\pi/2$.
Solid red (dark gray) and dashed black lines correspond respectively to $w^{\prime}=0.1, 0$. 
 Thick and thin lines correspond to $\Omega_0= 0.05, 0.3$, respectively.
An scheme of the setup is also included. }
\label{fig1}
\end{figure}

In this work we consider the setup of Fig. \ref{fig1}, which corresponds to a
double-barrier structure embedded into a ring. 
The ring is connected to two reservoirs with
the same chemical potential $\mu$ and two harmonic potentials with a phase-lag are applied at
the barriers.
Our aim is to perform a careful analysis
of the relevant symmetries of the system on the basis of the scheme of Refs. \cite{ser}
proposed for ratchet systems.   We investigate the possibility of a directed current in
the limit where the coupling to the reservoirs tends to zero as well as
 the role of the environment in the rectification properties of quantum pumps.

{\em Model}. 
We consider  a  tight-binding chain of $N$ sites
 with constant hopping $w$ and two barriers of height $E_b$. 
The ends of the chain
are  connected to reservoirs. The chain is closed with a hopping
 $w^{\prime}$ along the bond $\langle 1 N \rangle$. 
The Hamiltonian for the full system reads
\begin{eqnarray}
H &=& H_1+H_2 + H_{C}(t) - w_0 \sum_{k_1}(c^{\dagger}_1 c_{k_1} + H.c.) \nonumber\\
& & -  w_0 \sum_{k_2}(c^{\dagger}_N c_{k_2} + H.c.),
\end{eqnarray}
where $H_1$ and $H_2$ are free-electron Hamiltonians with degrees of freedom
labeled by $k_1, k_2$, respectively, which represent the reservoirs.
The latter 
are 
coupled to the ring through a hopping $w_0$. 
The Hamiltonian of the ring containing the two oscillating barriers reads
\begin{eqnarray}
& &H_C(t) = -w \sum_{l=1}^{N-1} (c^{\dagger}_l c_{l+1} + H.c.) -w^{\prime}
(c^{\dagger}_1 c_N + H.c.)+ \nonumber\\
& &  [E_b+V_1 \cos(\Omega_0 t + \delta) ] c^{\dagger}_{l_1} c_{l_1} 
+[E_b+V_2 \cos(\Omega_0 t ) ] c^{\dagger}_{l_2} c_{l_2}.
\end{eqnarray} 
This model has two interesting limits: for $w^{\prime}=0, w_0\neq 0$ it
corresponds to the linear array studied in Ref. \cite{lilip}, while
for  $w_0=0, w^{\prime}\neq 0$ it corresponds to the ring isolated from the reservoirs.
In the latter case, the spacial coordinates satisfy periodic boundary
conditions $l+N \equiv l$.
We assume that reservoirs and barriers are symmetrically placed
defining a mirror line along a diameter (see scheme of Fig \ref{fig1}).
Their positions satisfy $l_1=-l_2+1$. The relevant symmetry operations to analyze are:
\begin{eqnarray}
{\cal S}_1: & & l \rightarrow -l+1 , \;\;\;\;\; t \rightarrow t -
\delta/\Omega_0 \nonumber \\
{\cal S}_2: & & t \rightarrow -t ,
\end{eqnarray}
which cause a spacial inversion combined with a shift in the time coordinate, 
and
 a time inversion, respectively.

{\em Isolated annular pump.}
The ring isolated from the reservoirs ($w_0=0$) defines  a Hamiltonian,
or closed,
system. We now show that if $H_C(t)$ is invariant under ${\cal S}_1$ or
${\cal S}_2$, the directed current along the ring vanishes.

The evolution of a single particle wave function
$\Psi(t)= \sum_l \psi_l(t)$ is determined by (we work in units where
$\hbar=1$):
\begin{equation}
-i \frac{\partial}{\partial t} \psi_{l}(t) - 
\sum_{m=1}^{N} \varepsilon_{l,m}(t)   \psi_{m}(t) = 0,
\label{schro}
\end{equation}
being $\varepsilon_{l,m}(t) = - w_l \delta_{m,l+1} - w_{l-1} \delta_{m,l-1} 
+ \delta_{l,m} [ \delta_{l,l_1} v_1(t) +  \delta_{l,l_2}
v_2(t) ]$, being $w_l=w, l=1,\ldots,N-1$, $w_N=w^{\prime}$ and
$v_1(t)=E_b+V_1 \cos(\Omega_0 t + \delta)$, 
$v_2(t)=E_b+V_2 \cos(\Omega_0 t)$. 
The ensuing time-dependent current is 
\begin{equation}
J^{isol}_l(t)= e w_l \mbox{Im}[\psi^*_l(t) \psi_{l+1}(t) ].
\end{equation} 
Since the applied fields are harmonic with frequency $\Omega_0$, it is
verified $J^{isol}_l(t+\tau_0)=J^{isol}_l(t)$, being $\tau_0= 2 \pi / \Omega_0$.
The dc-component of this current is independent of $l$, due to the
continuity condition. It is defined as
\begin{equation}
J^{isol}= \frac{1}{\tau_0} \int_0^{\tau_0} J^{isol}_l(t) =
\frac{1}{N \tau_0} \sum_{l=1}^{N} \int_0^{\tau_0} J^{isol}_l(t).
\end{equation}

For $V_1=V_2$ and $\delta=0,\pi$, the matrix elements of the Hamiltonian 
$\varepsilon_{l,m}(t)$ are invariant under ${\cal S}_1$ and  ${\cal S}_2$.
Hence, applying ${\cal S}_1$ to the equations
(\ref{schro}), it is found $\psi_l(t) = \psi_{-l+1}( t -\delta/\Omega_0 )$,
while applying ${\cal S}_2$, it is found $\psi_l(t)=\psi_l^* (-t)$.
In the first case, we obtain $ J^{isol}_l(t) \rightarrow  - J^{isol}_{-l} (t
-\delta/\Omega_0 )$, while in the second one,  $J_l^{isol}(t) \rightarrow
-J_l^{isol}(-t)$. In the two cases, the final consequence is $J^{isol}=0$.

In summary, it becomes clear that, in order to have $J^{isol} \neq 0$
we need geometrical arrangements with broken ${\cal S}_1$ and ${\cal S}_2$ symmetries. 
In the case of a symmetric static setup like the one we are considering,
${\cal S}_1$ can be dynamically broken by applying time-dependent potentials with (i) $V_1 \neq V_2$
and/or (ii) $\delta \neq 0, \pi$. Instead,  condition (ii) must be fulfilled in order  to break ${\cal S}_2$.
Altogether, we conclude that in the Hamiltonian limit,  $\delta \neq 0, \pi$
 is a necessary condition that the pumped two barrier system
must fulfill in order to support a finite net current.  

{\em The presence of the environment.} We now refer to the setup of Fig. \ref{fig1}
with $w_0 \neq 0$. As mentioned before, the linear arrangement is contained in
the limiting case $w^{\prime}=0$. In the latter limit the dc-current vanishes as 
$w_0 \rightarrow 0$ \cite{lilip}. Instead, the above symmetry analysis suggests that
this may be not the case in the ring geometry ($w^{\prime} \neq 0$) when $\delta \neq 0, \pi$. 
A convenient theoretical framework to study transport
phenomena in driven open quantum systems is provided by non-equilibrium Green function 
formalism \cite{lilip,lilir}. 
For reservoirs with the same chemical potential $\mu$, 
the dc-component of the current flowing along the ring reads:
\begin{equation}
J_l= e \int_{-\infty}^{\infty}  d\omega f(\omega) T_l(\omega),
\label{trans}
\end{equation}
where we assume zero temperature, hence, $f(\omega)=\Theta(\mu -\omega)$.
The transmission function is \cite{lilip}:
\begin{eqnarray}
 T_l(\omega) & = &   \frac{1}{\pi \tau_0} w_l |w_0|^2 \int_0^{\tau_0} dt
 \rho_0(\omega) \mbox{Im}[
G^R_{l,1}(t,\omega) G^A_{1,l+1}(\omega,t)  \nonumber \\
& &+ G^R_{l,N}(t,\omega) G^A_{N,l+1}(\omega,t)] ,
\label{tl}
\end{eqnarray}
where $G^R_{l,m}(t,\omega)$ is the Fourier transform of the retarded Green
 function, with respect to the difference of time $t-t^{\prime}$,
 at the time of
observation $t$. The advanced Green function is
$G^A_{l,m}(\omega,t)=[G^R_{m,l}(t,\omega)]^*$ and $\rho_0(\omega)$ is the density of states of the reservoirs.
In this geometrical arrangement, there is in general a net charge flow between ring and reservoirs. Hence,
$T_l(\omega)=T(\omega)$, $J_l=J$ for $l=1,\ldots,N-1$ and
$T_N(\omega)=T^{\prime}(\omega)$, $J_N=J^{\prime}$, being $J^{\prime}\neq J$, which 
of course verify Kirchoff rules. The exact retarded Green function is the solution of the following 
linear set \cite{lilip}:
\begin{eqnarray}
& & G^R_{m,n}(t,\omega)  = 
G^0_{m,n}(\omega) \nonumber \\
& &+ \sum_{j=1}^2 \frac{V_j}{2} e^{i (\delta_j+\Omega_0 t)}  G^R_{m,l_j}(t,\omega-\Omega_0) 
G^0_{l_j,n}(\omega) \nonumber \\ 
& &+ 
 \sum_{j=1}^2 \frac{V_j}{2} e^{-i (\delta_j+\Omega_0 t)}  
G^R_{m,l_j}(t,\omega+\Omega_0) G^0_{l_j,n}(\omega),
\label{dyret}
\end{eqnarray}
where $\delta_1=\delta, \delta_2=0$, while 
$G^0_{m,n}(\omega)$ is the retarded Green function of the equilibrium ring
with barriers connected to reservoirs
without time-dependent voltages. 

We now show results for different pumping conditions and strengths of coupling to the reservoirs. 
We consider two infinite tight-binding chains of bandwidth $W$ for the reservoirs, described by the
density of states $\rho_0(\omega)=4 \sqrt{1-\omega^2/W^2} \Theta(W - \omega)$. 
All energies
are expressed in units of $w$. 

First, we would like to point out that the coupling to the reservoirs introduce
inelastic scattering events and the propagation of the wave-packet along the
ring looses its coherence. Hence, for strong coupling to the reservoirs, the system
evolves smoothly from the annular to the linear geometry as $w^{\prime}
\rightarrow 0$ and we expect that the transport properties of the ring does not significantly
differ from those of the linear array. 
This feature is illustrated in Fig. \ref{fig1}, where we show that for $w_0=1$ and
small $w^{\prime}$ we can almost exactly reproduce the transmission function $T(\omega)$ 
of the linear array. 

\begin{figure}
\centerline{\psfig{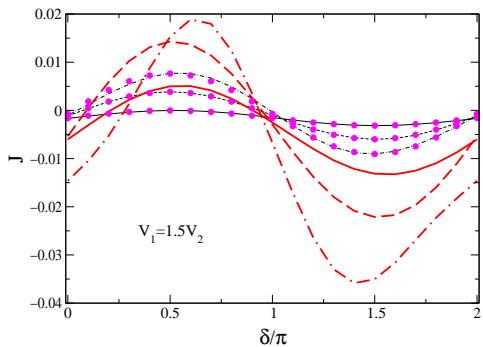}}
\hfill
\caption{\small (Color online) dc- current $J$ as a function of the phase lag for
 $V_2=1.5 V_2$ and $w_0^2=0.5$. Thin black and thick red (dark gray)
 lines correspond to $V_1=0.2,
 0.5$, respectively. Solid, dashed and dot-dashed lines correspond to
$\Omega_0= 0.05,0.1,0.6$. Circles correspond to fits with the function
$J=A_0+A_1 \sin(\delta)$.
 The chemical potential
is $\mu=0.1$ and  $w^{\prime}=1$. Other
parameters are as in Fig. \ref{fig1}.} 
\label{fig2}
\end{figure}

In what follows, we consider $w^{\prime}=w$ and
a symmetric static arrangement with barrier positions $l_1$
and $l_2$ equidistant to the reservoirs ($l_1=-l_2+1$). We
turn to show that a finite net current 
$J$ may be obtained for $\delta=0,\pi$
if the pumping amplitudes are different ($V_1 \neq V_2$). For small $V_1,V_2$, eq.
(\ref{dyret}) can be solved perturbatively and when this solution is replaced
in (\ref{tl}) for $l=N/2$, 
it is obtained
\begin{eqnarray}
& & T(\omega)  \sim   w |w_0|^2 \rho_0(\omega) \{
[(\frac{V_1}{2})^2-(\frac{V_2}{2})^2 ]
\gamma_1(\omega) \nonumber \\
& & \times \mbox{Im}[G^0_{N/2,1}(\omega+ \Omega_0 ){G^0}^*_{N/2+1,1}(\omega+ \Omega_0 ) + \nonumber \\
& & G^0_{N/2,1}(\omega- \Omega_0 ) {G^0}^*_{N/2+1,1}(\omega- \Omega_0 ) ] \nonumber \\
& &+ \frac{V_1 V_2}{2} \sin(\delta) \; \gamma_2(\omega) \times \nonumber \\
& & [ |G^0_{N/2,1} (\omega+ \Omega_0 )|^2-|G^0_{N/2,N} (\omega+ \Omega_0 )|^2\nonumber \\
& &
-|G^0_{N/2,1} (\omega- \Omega_0 )|^2 + |G^0_{N/2,N} (\omega- \Omega_0 )|^2]
\},
\label{tlo}
\end{eqnarray}
being $\gamma_1(\omega)=|G^0_{l_1,1} (\omega)|^2+ |G^0_{l_1,N} (\omega)|^2$
and $\gamma_2(\omega)=\mbox{Re}[G^0_{l_1,1} (\omega)[G^0_{l_1,N}
 (\omega)]^*]$.
Therefore, it is found that the net current behaves as $J \sim B_0 [(\frac{V_1}{2})^2-(\frac{V_2}{2})^2 ] 
+ B_1 V_1 V_2 \sin(\delta) $, being $B_0, B_1$ real coefficients.  The exact solution of $J$ as 
a function of $\delta$  is shown in Fig. \ref{fig2}.
For the smallest $V_1$, there is agreement with the functional behavior
suggested by (\ref{tlo}). For higher $V_1$, departures from this
behavior are observed. In any case, the  feature we want to emphasize
is  that for $\delta=0, \pi$, symmetry ${\cal S}_1$ is dynamically broken for $V_1 \neq V_2$,
but ${\cal S}_2$ is still an exact symmetry of the Hamiltonian $H_C(t)$ for
the isolated system. 
A non-vanishing $J$ at these points is a consequence of the fact that the latter symmetry is
broken due to the coupling to the reservoirs. Another issue worth mentioning is that
the functional behavior of $J$ we are finding is just the one observed in the experimental
work \cite{swi}.

\begin{figure}
\centerline{\psfig{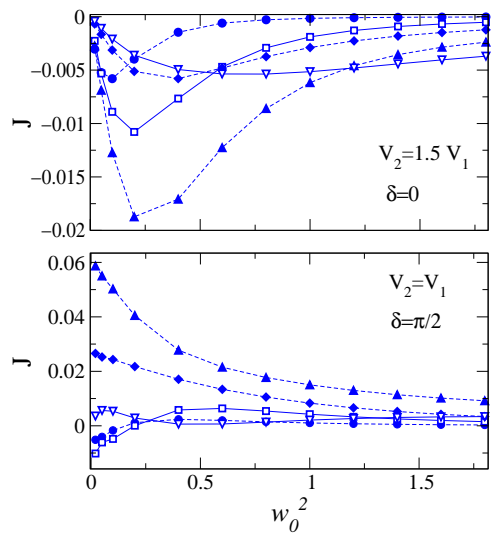}}
\hfill
\caption{\small (Color online) dc- current $J$ as a function of the coupling to the
reservoirs. Upper panel corresponds to $V_1=1.5 V_2$ and $\delta=0$ and
lower panel corresponds to $V_1=V_2$, being $V_1=0.5$ and $\delta=\pi/2$. Circles, squares,
diamonds, upper and lower triangles correspond to
$\Omega_0=0.01,0.05,0.45,0.6,0.75$, respectively.Other parameters are as in Fig. \ref{fig2}.  } 
\label{fig3}
\end{figure}

The role of the environment in breaking time-reversal symmetry is highlighted in Fig.
\ref{fig3}. For finite $w_0$, ${\cal S}_2$ is broken due to the coupling to
the reservoirs. The figure illustrates the behavior for  
${\cal S}_1$  dynamically broken in two different ways:
The upper panel corresponds to $\delta=0$ and $V_1 \neq V_2$, while the lower one, to
$V_1=V_2$ and $\delta=\pi/2$. In the first case 
symmetry ${\cal S}_2$ is restored as $w_0 \rightarrow 0$. Thus, $J \rightarrow 0$ as
the system evolves towards the Hamiltonian limit. Instead in the second case, ${\cal S}_2$ 
remains broken at $w_0=0$ and $J$ may achieve a finite value. 

\begin{figure}
\centerline{\psfig{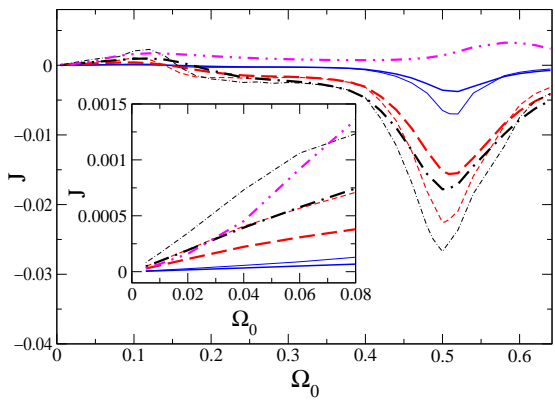}}
\hfill
\caption{\small (Color online) dc-current $J$ as a function of the pumping frequency 
$\Omega_0$ for $\delta=\pi/2$. Thin and thick lines correspond to couplings to the reservoirs 
$w_0^2=0.5, 1$, respectively. Blue (black) solid and red (dark grey) dashed lines correspond to
$\mu=-1.14$ and
$V_1=V_2=0.2,0.5$, respectively. Black dashed-dotted lines correspond to
$\mu=-1.14$ and
$V_2=1.5 V_1$ with $V_1=0.5$. Magenta (gray) two dots-dashed line corresponds to
$V_1=V_2=0.2$ and $\mu=0.1$.
Other parameters are as in Fig. \ref{fig2}.
The inset shows a zoom for small $\Omega_0$.
} 
\label{fig4}
\end{figure}
Finally, let us comment on the behavior of $J$ as a function of the pumping frequency 
$\Omega_0$. For low $V_1, V_2$, an expansion of (\ref{tlo}) in powers of  $\Omega_0$,
leads to $J \propto \Omega_0$.  For completeness, we also show in Fig. \ref{fig4} the exact behavior
of $J$ for arbitrary $\Omega_0$, $V_1$ and $V_2$. A large $|J|$ is obtained when
$\Omega_0$ is resonant (i.e when it coincides with the energy difference
between two levels of the isolated system). 
The inset shows that $J$ changes linearly in $\Omega_0$ 
for small enough pumping frequencies, which is typical of adiabatic driving
\cite{qupumpt,mosk1}. Such a linear behavior is, however, not expected when the coupling to
the environment vanishes. This is because in that limit $G^0_{m,n}(\omega)$ corresponds to a 
sequence of poles at the energy levels of the free ring and it is not possible to
perform a power expansion of 
(\ref{tlo}). Furthermore, for small pumping amplitudes, the structure of 
(\ref{dyret}) suggests that $J$ is only sizable for resonant $\Omega_0$, in agreement with 
Ref. \cite{mosk2}.

{\em Discussion}. 
Previous discussions in the literature on the role of symmetries 
in quantum pumps suggest different  behavior in  adiabatic and non-adiabatic regimes.
The idea of adiabatic  driving is associated to
small pumping amplitudes and low frequency $\Omega_0$ compared to the inverse
of the typical time for the particle propagation through the device. 
While in
practice this definition implies $J \propto \Omega_0$, this concept is usually formulated in terms of
some approximation for the scattering matrix \cite{qupumpt,mosk1} and it has
been pointed out   that such definition  strictly applies only to
isolated systems \cite{mosk2}. 

The symmetry analysis carried out in this work leads us to conclude that in open quantum systems
the fundamental condition to be fulfilled, in order to obtain a dc current, is spacial inversion
symmetry breaking in the Hamiltonian $H_C(t)$. Since we have not introduced any assumption
regarding the pumping amplitudes and frequencies, we conclude that this condition should
apply to both adiabatic and non-adiabatic regimes. 
Let us support with examples the fact that this conclusion applies, in particular, to the adiabatic regime. We
can mention, at least, three models with time inversion invariance in 
the Hamiltonian  limit
but broken spacial inversion symmetry
where $J$ behaves linearly in the pumping frequency:
(i) The system considered in the present work with $\delta=0,\pi$,
$V_1 \neq V_2$, finite $w_0$
and arbitrary (even zero) $w^{\prime}$ (see Fig. \ref{fig4}). 
(ii) Only one pumping potential applied away from a 
symmetric point under spacial inversion. 
In \cite{lilip} this problem has been
solved in a linear array and identical procedure and conclusions apply for the annular
geometry with contacts to reservoirs. (iii) A ring
threaded by a linear time-dependent 
flux coupled to reservoirs. This basic problem 
generated interesting discussions some time ago on the nature of resistive behavior \cite{ringo}.
More recently it has been 
 exactly solved \cite{lilir}. For low enough driving, the dc current is linear in the
induced emf, which is the effective pumping frequency of this problem.
In all these systems, the key point is that time-inversion symmetry breaking
is introduced by 
their coupling to the environment. Another important
conclusion is that
time-inversion symmetry breaking in the Hamiltonian $H_C(t)$ is, instead
 necessary to obtain a dc-current
in the isolated ring. In order to achieve this,  a minimum of two
time-dependent voltages with a phase lag non-commensurate with $\pi$ are needed. 

To finalize,
 we mention that the setup of Fig \ref{fig1} with vanishing $w^{\prime}$ and
finite $w_0$ can be viewed
as a schematic model to capture the role of symmetries in the transport
properties  of  the system studied in  Ref. \cite{swi}. Our results
are consistent with 
the behavior $J=A+B\sin(\delta)$ observed in that
experiment and suggest that the small finite $J$ at $\delta=0,\pi$ 
can be naturally explained as a consequence
of a slight difference in the amplitudes of the pumping voltages.

{\em Acknowledgments}. 
The author thanks S. Flach, S. Denysov and V. Gopar for
useful conversations, as well as the hospitality of D. Zanchi at
LPTHE-Jussieu-Paris. 
 Support from 
PICT 03-11609 from Argentina, BFM2003-08532-C02-01
from MCEyC of Spain, grant ``Grupo consolidado DGA''
and from the MCEyC of Spain through ``Ramon y Cajal'' program 
 are
acknowledged. LA is staff member of CONICET, Argentina.
Simulations were performed at BIFI cluster.

\end{document}